\documentclass[apj]{emulateapj}

\shorttitle{The high jet efficiency in FRII}

\shortauthors{S.-L. Li}

\begin{document}

\title{The high efficiency jets magnetically accelerated from a thin disk in powerful lobe-dominated FRII radio galaxies}

\author{Shuang-Liang Li\altaffilmark{1,2}}


\altaffiltext{1}{Key Laboratory for Research in Galaxies and
Cosmology, Shanghai Astronomical Observatory, Chinese Academy of
Sciences, 80 Nandan Road, Shanghai, 200030, China; lisl@shao.ac.cn}
\altaffiltext{2}{JILA, University of Colorado and National Institute of Standards and Technology, 440 UCB, Boulder, CO 80309, USA}

\begin{abstract}

A maximum jet efficiency line $R\sim 25$ ($R=L_{\rm jet}/L_{\rm
bol}$), found in FRII radio galaxies by Fernandes et al., was
extended to cover the full range of jet power by Punsly. Recent general relativistic magnetohydrodynamic (GRMHD) simulations on jet formation mainly focused on the enhancement of jet power. In this work, we suggest that the jet efficiency could be very high even for conventional jet power if the radiative efficiency of disk were much smaller. We adopt the model of a thin disk with magnetically driven winds to investigate the observational high efficiency jets in
FRII radio galaxies. It is found that the structure of a thin disk
can be significantly altered by the feedback of
winds. The temperature of disk gradually decreases
with increasing magnetic field; the disk density, surface
density and pressure also change enormously. The lower temperature and higher
surface density in inner disk result in the rapid decrease of
radiative efficiency. Thus, the jet efficiency is greatly improved even the jet power is conventional. Our results can
explain the observations quite well. A theoretical maximum jet efficiency $R \sim 1000$ suggested
by our calculations is large enough to explain all the high jet efficiency
in observations even considering the episodic activity of jets.

\end{abstract}


\section{INTRODUCTION}

Relativistic jets are common characters of active galactic nuclei
(AGNs). According to the different morphology of their radio
structures, the radio galaxies can be divided into two classes,
i.e., FRI (defined by edge-darkened radio lobes) and FRII (defined
by edge-brightened radio lobes and hot spots) \citep{f1974}. Several
recent observations discovered that the jet efficiency (defined as
$R=L_{\rm jet}/L_{\rm bol}$) of some luminous lobe-dominated FRII
could be very high, where $L_{\rm jet}$ and $L_{\rm bol}$ are the
jet power and the bolometric luminosity of AGNs, respectively \citep*[][hereafer P11]{m2011,f2011,p2011}. A
maximum jet efficiency $R \sim 25$ was found by
\citet{f2011}, where they adopted a complete
sample of the most powerful radio galaxies at redshift $z\sim1$.
P11 extended the results of \citet{f2011} and found that the maximum jet efficiency
line can cover the full four order of magnitudes of jet power by including other samples, such as, a
blazar sample in \citet{g2010}, a small sample of X-ray cavities in
\citet{m2011}, etc.

There are mainly two most popular jet formation mechanisms so far,
i.e., the Blandford-Znajek (BZ) process \citep{b1977} and the
Blandford-Payne (BP) process \citep{b1982}. In BZ process, the
rotating energy of a black hole can be extracted to power a jet by
the large scale magnetic fields maintained by an accretion disk. But in
BP process, the jet power comes from the rotating disk itself
instead. The bolometric luminosity of a radio galaxy can be
expressed as $L_{\rm bol}=\eta_{\rm th} \dot{M}c^2$, where
$\eta_{\rm th}$ is the radiative efficiency of the accretion disk.
If we specify the jet power as a function of accretion rate, $L_{\rm
jet}=\eta_{\rm Q} \dot{M}c^2$ (where $\eta_{\rm Q}$ is the jet production efficiency), the jet efficiency $R(=L_{\rm
jet}/L_{\rm bol}=\eta_{\rm Q}/\eta_{\rm th})$ will be decided by both
$\eta_{\rm Q}$ and $\eta_{\rm th}$. The obviously different accretion
rate between FRI and FRII radio galaxies implies that they should
have different accretion model \citep{l1996,g2001,x2009}. In the
general picture of FRII radio galaxies, jet is supposed to be
launched from a ridiatively efficient accretion disk , where
$\eta_{\rm th}$ varies from about $0.06$ to $0.4$ for a non-rotating
black hole and a extreme Kerr black hole, respectively. Recent GRMHD simulations on jet formation mainly focused on the improvement of jet power
$L_{\rm jet}$ in order to explain the observed high jet production efficiency $\eta_{\rm Q}$. The jet production efficiency for a magnetically-arrested-disk (MAD) can reach $\sim 30\%$ and $140\%$ for $a=0.5$ and $0.99$, respectively \citep{t2011}, which implies that the black hole spin parameter $a$ may play a key role in the
formation of jets \citep{m2005,t2010,t2011}. However, the observed $\eta_{\rm Q}$ is based on an assumption that the radiative efficiency $\eta_{\rm th}\sim 0.1$, which is suggested from observational
constraints on the growth of massive black holes
\citep*[e.g.,][]{y2002}. Thus, if the radiative efficiency of a disk were much smaller than $0.1$, it is possible to get a high jet efficiency $R$ even for conventional jet production efficiency $\eta_{\rm Q}$.

\citet{l2012} and \citet{l2014} do seem to provide a possible way for this picture.
They produced a model for a thin disk with magnetically driven
winds/jets, in which the angular momentum and energy carried away
by jets are properly included. It was found that the disk properties
can be changed significantly by the feedback of jets. For example,
the temperature of a thin disk with jets is obviously lower
compared with that of a standard thin disk \citep*[see figure 1,][]{l2012}
for the reason that a large fraction of gravitational
energy released in the disk is carried away by jets. This will
directly result in the decrease of bolometric luminosity (and
$\eta_{\rm th}$) of the accretion disk and the increase of jet
efficiency $R$. In this work, we will detailedly investigate the structure of a thin disk with winds and the
high efficiency jets in powerful lobe-dominated FRII radio galaxies.


\section{MODEL}\label{models}

We consider a thin disk with magnetically driven winds surrounding a spinning black hole. The continuity, radial momentum, angular momentum and energy equations of the disk are as follows:
\begin{equation}
\frac{d}{dr}(2\pi \Delta^{1/2} \Sigma v_{\rm r})+4\pi r \dot{m}_{\rm w}=0, \label{continuity}
\end{equation}

\begin{equation}
\frac{\gamma_\phi A M}{r^4 \Delta}\frac{(\Omega-\Omega_{\rm k}
^+)(\Omega-\Omega_{\rm k} ^-)}{\Omega_{\rm k} ^+ \Omega_{\rm k}
^-}+g_{\rm m}=0, \label{momentum}
\end{equation}

\begin{equation}
-\frac{\dot{M}}{2\pi} \frac{dL}{dr} + \frac{d}{dr}(r
W^r_\phi)+T_{\rm m}r=0, \label{angular}
\end{equation}

\begin{equation}
\nu \Sigma \frac{\gamma_{\phi}^4 A^2}{r^6}\left(
\frac{d\Omega}{dr}\right)^2= \frac{16acT^4}{3\bar{\kappa}\Sigma},
\label{energy}
\end{equation}
where the equations and the meaning of parameters are all the same as \cite{l2014}. We focus on the fast moving jets from the disk and adopt $B_\phi = 0.1B_{\rm p}$, which corresponds to the case of fast moving jets with low mass-loss rate
\citep{o1998,c2002b,l2014}. So we can simply ignore the mass loss rate term in the continuity equation.

The bolometric luminosity of accretion disk can be
calculated as
\begin{equation}
L_{\rm bol}=\int^{r_{\rm out}}_{r_{\rm in}} Q_{\rm rad}2 \pi r
dr=\eta_{\rm th} \dot{M} c^2,
\end{equation}
where $r_{\rm in}=r_{\rm isco}$ and $r_{\rm out}=1000 r_{\rm g}$ ($r_{\rm g}=GM/c^2$) are
the inner and outer radius of disk, respectively, $r_{\rm isco}$ is
the innermost stable circular orbit of disk. $Q_{\rm
rad}={16acT^4}/{3\bar{\kappa}\Sigma}$ is the radiative cooling rate
per unit surface and $\eta_{\rm th}$ is the radiative efficiency of
the accretion disk.

Jets can be powered by both BZ and BP processes. In the general
form of the BZ process, the jet power $L_{\rm BZ}$ can be estimated
with \citep{g1997}
\begin{equation}
L_{\rm BZ}=\frac{1}{32}\omega^{2}_{\rm F}B^2_\perp r^2_{\rm H}
(J/J_{\rm max})^2 c. \label{lbz}
\end{equation}
where $r_{\rm H}$ is the horizon radius, $B_\perp$ is the component
of the magnetic field normal to the black hole horizon, $J$ and
$J_{\rm max}=GM^2/c$ are the angular momentum and maximum angular
momentum of a black hole, and $\omega_{\rm F}^2\equiv \Omega_{\rm F}
(\Omega_{\rm H}-\Omega_{\rm F})/\Omega_{\rm H}^2$ is a factor at
black hole horizon determined by the angular velocity of black hole
and magnetic filed lines. The strength of the field
threading the horizon of black hole ($B_\perp$) is comparable to that
threading the inner region of the disk \citep{l1999}. In this work, we
adopt $B_\perp$ as the maximal magnetic field strength in the disk
for simplicity. The power of the jets accelerated from an accretion disk (BP
process) can be calculated with \citep*[e.g.,][]{l1999,c2002}
\begin{equation}
L_{\rm BP}=\int_{r_{\rm in}}^{r_{\rm out}} \frac{B_{\rm p}B_{\rm
\phi}}{4\pi} r \Omega 2\pi r {\rm d} r, \label{lbp}
\end{equation}
where $B_{\phi}$ and $B_{\rm p}$ are the toroidal and poloidal components of magnetic field, respectively.

Thus the total jet power $L_{\rm jet}$ is:
\begin{equation}
L_{\rm jet}=L_{\rm BZ}+L_{\rm BP}=\eta_{\rm Q} \dot{M} c^2.
\label{ljet}
\end{equation}
With $\eta_{\rm th}$ and $\eta_{\rm Q}$, the jet efficiency is
given by:

\begin{equation}
R=\frac{\eta_{\rm Q}}{\eta_{\rm th}}. \label{lbz}
\end{equation}

\section{RESULTS}\label{results}

\begin{figure*}
\includegraphics[width=16cm]{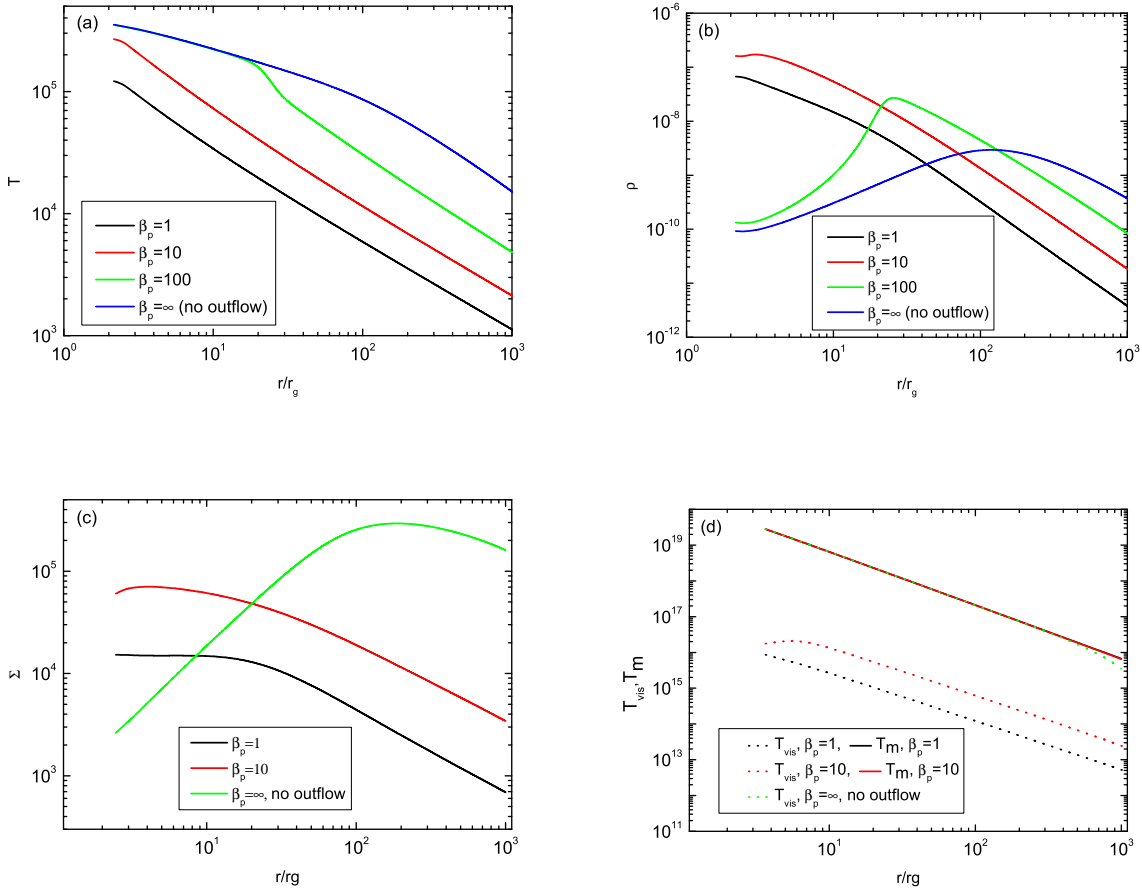}
\caption{The disk properties of a relativistic thin accretion disk
with magnetically driven outflows/jets, where $a=0.9$ is adopted.
(a),(b) represent the disk temperature and density for different
$\beta_{\rm p}$, respectively. From the bottom up, the black, red, green and
blue lines are for $\beta_{\rm p}=1, 10, 100$ and $\infty$, respectively. (c)
represents the disk surface density for different $\beta_{\rm p}$, the
black, red and green lines are for $\beta_{\rm p}=1, 10$ and $\infty$,
respectively. (d) represents the viscous torque and magnetic torque
for $\beta_{\rm p}=1, 10$ and $\infty$, respectively. The solid and dotted
lines are for magnetic torque and viscous torque, respectively. The
black, red and green lines are for $\beta_{\rm p}=1, 10$ and $\infty$,
respectively. \label{properties}}
\end{figure*}

\begin{figure*}
\includegraphics[width=16cm]{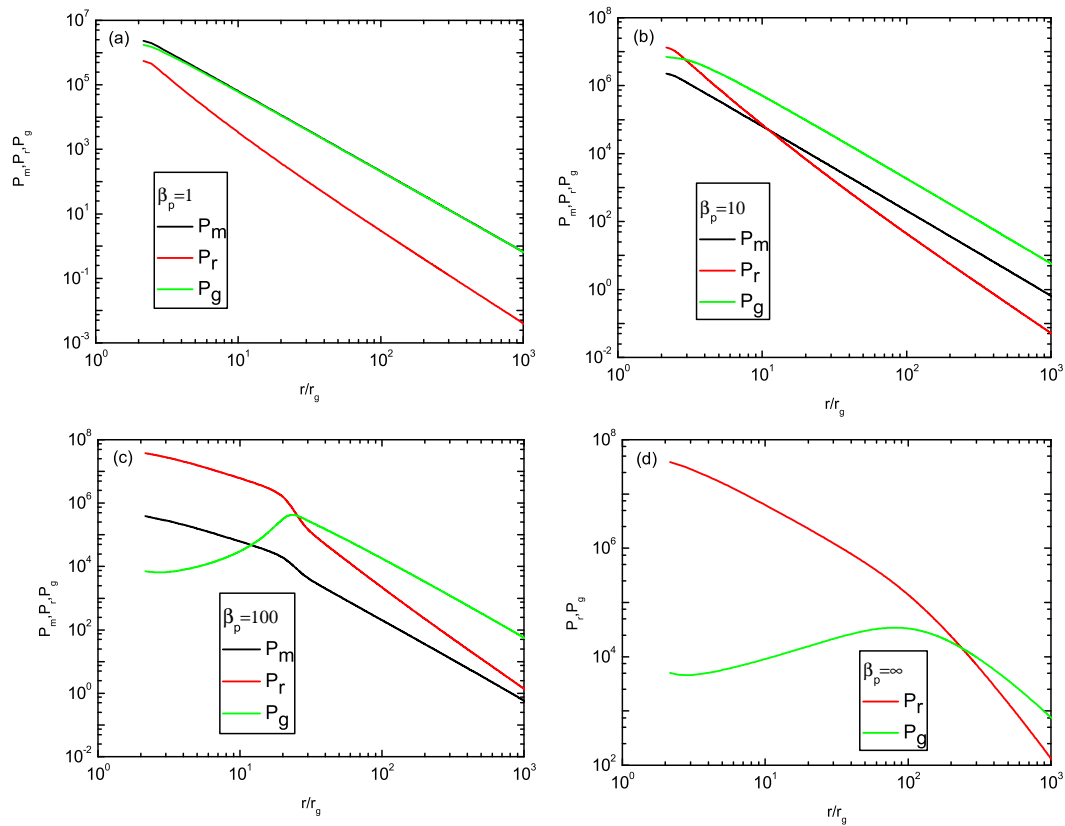}
\caption{The disk pressures as functions of radius, i.e., the gas,
radiation and magnetic pressure of a disk for $\beta_{\rm p}=1, 10, 100$
and $\infty$, respectively, where $a=0.9$ is adopted. The black, red and
green lines are for magnetic pressure, radiation pressure and gas
pressure, respectively. \label{pressure}}
\end{figure*}

\begin{figure}
\includegraphics[width=8cm]{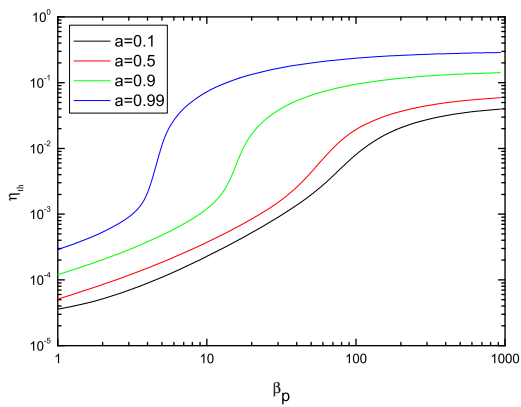}
\caption{Radiative efficiency of a relativistic thin accretion disk
with magnetically driven jets as functions of $\beta_{\rm p}$ for different
spin $a$. From the bottom up, the black, red, green and blue lines
are for $a=0.1, 0.5, 0.9$ and $0.99$, respectively.
\label{radiative}}
\end{figure}

\begin{figure}
\includegraphics[width=8cm]{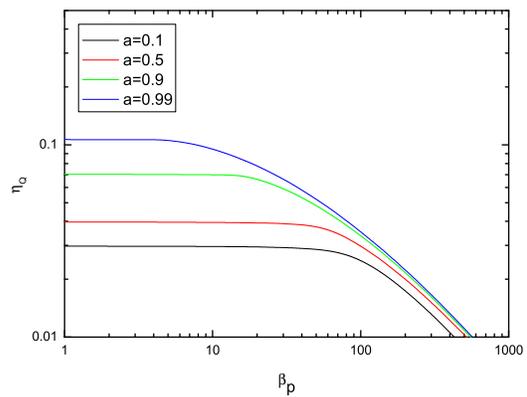}
\caption{Same as Figure \ref{radiative}, except that this figure
represent the jet production efficiency. \label{jet}}
\end{figure}

\begin{figure}
\includegraphics[width=8cm]{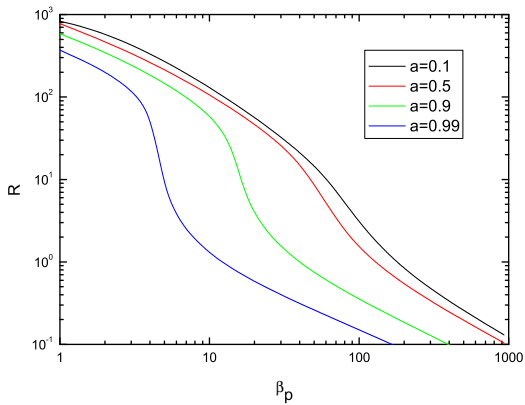}
\caption{Theoretical jet efficiency $R$ as functions of
$\beta_{\rm p}$ for different black hole spin. From the top down, the black,
red, green, and blue lines are for $a=0.1, 0.5, 0.9$ and $0.99$,
respectively. \label{r}}
\end{figure}

\begin{figure}
\includegraphics[width=8cm]{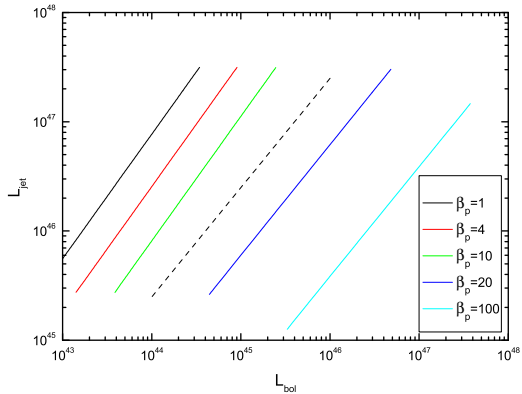}
\caption{Jet power as functions of bolometric luminosity for
different $\beta_{\rm p}$ with $a=0.9$, where the black hole mass varies
from $10^8 M_{\odot}$ to $10^{10} M_{\odot}$. The dashed
line corresponds to the observational jet efficiency
lines $R=25$. The black, red, green, blue
and cyan lines are for $\beta_{\rm p}=1, 4, 10, 20$ and $100$,
respectively. \label{ljet}}
\end{figure}

\begin{figure}
\includegraphics[width=8cm]{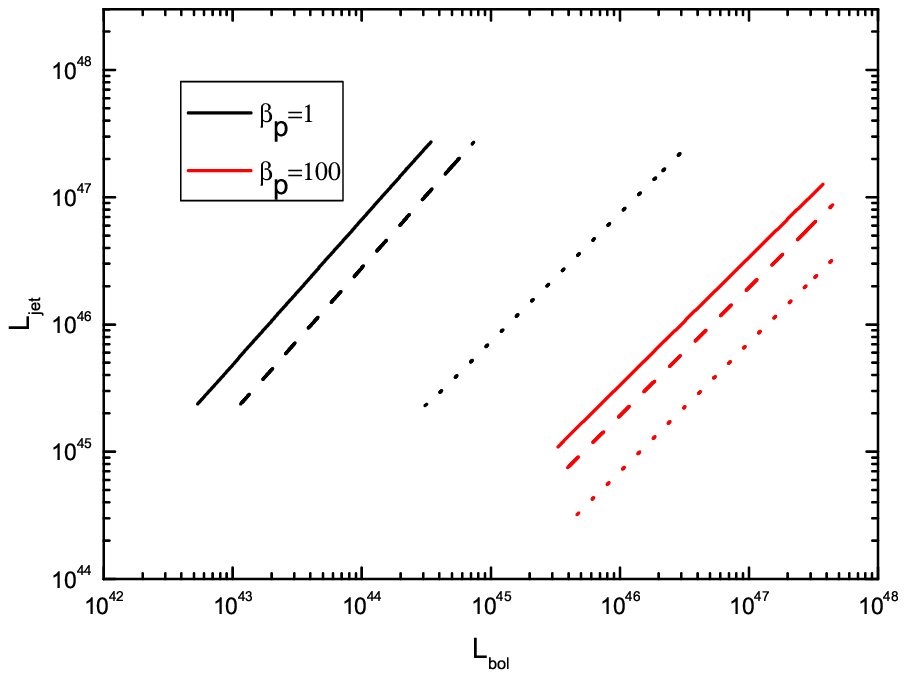}
\caption{Jet power as functions of bolometric luminosity for
different $\beta_{\rm p}$ and $\theta$ with $a=0.9$, where the black hole
mass varies from $10^8 M_{\odot}$ to $10^{10} M_{\odot}$. The black
and red lines are corresponding to $\beta_{\rm p}=1$ and $\beta_{\rm p}=100$,
respectively. The solid, dashed and dotted lines are for $\theta=60,
30$ and $10$ degrees, respectively. \label{ljettheta}}
\end{figure}

We solve equations
(\ref{continuity}) - (\ref{energy}) to achieve the structure of a thin disk with winds
by using the same numerical methods as \citet{l2014}.
In all the calculations, we adopt the conventional viscosity parameter $\alpha=0.1$, black
hole mass $m=M/M_\odot=10^9$ (except for figures \ref{ljet}, \ref{ljettheta}) and
Eddington-scale accretion rate $\dot{m}=\dot{M}/\dot{M}_{\rm
Edd}=0.25$, where $\dot{M}_{\rm Edd}=1.5\times10^{18}m \rm {g/s}$
and $\dot{m}=0.25$ is the typical value of broad-line AGNs
\citep{k2006}. Our results are qualitatively the same for different
black hole mass and accretion rate.

In figure \ref{properties}, it is found that the disk structure has
been significantly altered by the winds. The
temperature of a thin disk with winds is obviously lower than
that of a standard thin disk, because a large fraction of the
gravitational energy released in the disk is carried away by the
jets (see figure \ref{properties}a), and gradually decreases with
increasing magnetic field strength [smaller $\beta_{\rm p}$, $\beta_{\rm p}=(P_{\rm {gas}}+P_{\rm rad})/(B_{\rm p}^2/8\pi)$]. Except for the temperature, the disk density and surface density also exhibit
enormous changes as presented in figure \ref{properties}b, c.
Especially for the inner disk region, both the density and surface
density increase greatly. In figure \ref{properties}d, it is found
the magnetic torque is far more larger than the viscous torque for
the disk with strong winds. When the magnetic field is strong
enough ($\beta_{\rm p} \leq 10$), the radiative pressure dominated inner
region in a thin disk vanishes and the whole disk is
dominated by gas pressure (figure \ref{pressure}). The change of
pressure in inner disk also induces the change of disk density: the
initial negative correlation between the density and radius has
become positive (figure \ref{properties}b), which is a result of gas
pressure dominated inner disk.

The radiative efficiency $\eta_{\rm th}$ varied with $\beta_{\rm p}$ for
different black hole spin is plotted in figure \ref{radiative}. In this work, a minimal $\beta_{\rm p} \sim 1$ is adopted for the reasons that the jet efficiency $R$ ($10^{2-3}$, see figure \ref{r}) is high enough for $\beta_{\rm p} \sim 1$ and that MRI will be totally suppressed if $\beta_{\rm p}$ is smaller than $1$. $\eta_{\rm th}$ decreases fast with decreasing $\beta_{\rm p}$ and can be as
low as $10^{-3}-10^{-5}$ for different $a$ with $\beta_{\rm p}=1$, which is
about $0.1\%$ of the efficiency of a standard thin disk. Contrary to
$\eta_{\rm th}$, the jet production efficiency $\eta_{\rm Q}$ decreases
rapidly with the increase of $\beta_{\rm p}$ (figure \ref{jet}). The structure of a thin disk
with lower spin is found to be changed relatively easier. But
the results are qualitatively the same for different spin $a$ (figures, \ref{radiative} and \ref{jet}).

As $\beta_{\rm p}$ becomes smaller and smaller, the simultaneous increase of
$\eta_{\rm Q}$ and the decrease of $\eta_{\rm th}$ will result in a
rapid increase of jet efficiency $R$ (see figure \ref{r}). The
maximum theoretical jet efficiency can reach from several
hundreds to $1000$ for different spin $a$ with $\beta_{\rm p}=1$, which can
easily explain the observational high jet efficiency in luminous
FRII radio galaxies. In figure \ref{ljet}, we also draw the jet
power as functions of disk bolometric luminosity as figure $1$ in
P11, where the black hole mass varies from $10^8 M_{\odot}$ to
$10^{10} M_{\odot}$. Compared with the observational maximum jet
efficiency (the dashed line), our results suggest that the magnetic
pressure is about $5-10\%$ of the gas+radiation pressure in the disk
($10< \beta_{\rm p} <20$) when $\alpha=0.1$ and $B_{\phi}=0.1 B_{\rm p}$ are adopted.

An inclination angle $\theta$ of the field line with respect to the disk surface
is required to be smaller than $60$ degrees
in order to successfully launch jets from a cold thin disk \citep{b1982,c2012}. We simply
adopt an angle of $60$ degrees in this work except for figure
\ref{ljettheta}. In figure \ref{ljettheta}, we
investigate the effects of inclination angle $\theta$ on the jet power. It is found that the inclination
angle strongly affects the jet power and the jet
efficiency fast decreases with decreasing $\theta$. For $\beta_{\rm p}=1$,
the jet efficiency decreases for about two order of magnitudes when
$\theta$ varies from $60$ degrees to $10$ degrees. The jet efficiency
$R$ is reduced to $\sim 10$ for $\beta_{\rm p}=1$ and $\theta=10^{\circ}$.

\section{CONCLUSIONS AND DISCUSSION}\label{sec:summary}

The basic physics of the jet formation models adopted in this work
are the same as the previous works of \citet{g1997} and
\citet{l1999}. However, their estimates of the magnetic field
strength of a disk are based on the conventional accretion disk
models without magnetic field, which is a good approximation in the
weak magnetic field case as the disk structure has not been altered
significantly by the field. But that assumption becomes invalid if
the field is strong. The high power carried by the jets has greatly changed
the structure of a thin disk (figure \ref{properties},
\ref{pressure}). Except for the angular velocity, which is close to
Keplerian angular velocity for a thin disk \citep{l2014}, all other disk
properties (temperature, density, surface density, pressure) change
a lot compared with the standard disk. The magnetic torque $T_{\rm
m}$ is found to dominate over the viscous torque $T_{\rm vis}$ (see figure \ref{properties}d). The temperature decreases and the density increases significantly with decreasing $\beta_{\rm p}$. It is the increase of surface density and
the decrease of disk temperature in the inner disk region that
result in the very low radiative efficiency in the disk. The
obviously lower temperature compared with that of a standard thin disk
even leads to the disappearance of the inner disk dominated by
radiative pressure (figure \ref{pressure}). Thus, the thin disk becomes both thermally and viscously stable
on the presence of disk winds \citep{l2014}.

Previous efforts mainly focused on the improvement of jet power
$L_{\rm jet}$. But indeed, if the accumulation of magnetic flux in the inner region of
accretion disk is considered \citep{n2003,porth2010,porth2011,t2011,m2012,s2013}, not only can the jet production efficiency be improved significantly, the radiative efficiency can also decrease a lot. Thus, a very high jet production efficiency $\eta_{\rm Q}$ isn't always needed. The jet efficiency could be quite 'normal' if the radiative efficiency
decreases significantly (see figure \ref{radiative}, \ref{jet}).
According to our model, the reason for the high jet efficiency is
that most of released gravitational energy in the disk is carried
away by jets, which results in the very low radiative
efficiency. A theoretical maximum jet efficiency $R\sim 1000$ is
found, which is large enough even we take the episodic
activity of jets into account. Compared with the observational
results in P11 (the dashed line in figure \ref{ljet}), our study
indicates that $10< \beta_{\rm p}<20$ is required for $\alpha=0.1$ and $B_{\phi}=0.1 B_{\rm p}$.
But if we consider smaller $\alpha$ and larger proportion of $B_{\phi}$ in the
field, the magnetic torque could be more dominant \citep{l2014}. Thus the required field could be much
weaker (larger $\beta_{\rm p}$). From figure \ref{ljettheta}, it seems that the jet efficiency
has a positive relation with $\theta$. But if we consider the evolution of field in a thin disk, the
smaller $\theta$ may represent a stronger magnetic field \citep{l1994a,c2013}, which should thus help to improve the jet efficiency.

The large scale magnetic field plays a key role in the formation of jet. But how
it forms is still an unsolved problem. A popular mechanism is that the field lines can be dragged inwards
with the accretion of gas. The magnification of field seems to be hard in a thin disk because the speed of turbulent diffusion is faster than that of advection \citep{v1989,l1994a}. But when taking the magnetically driven winds into account \citep{c2013,l2014}, the field may be effectively magnified even for a thin disk with very weak initial field
($\beta_{\rm p}\sim 10^3$). However, a balance of magnetic field advection and diffusion is required in order to avoid the formation of a MAD. How such a balance can be achieved and kept stable are still unclear \citep*[e.g.,][]{c2002b,c2013,b2012}. \citet{l1994b} argued that a disk-wind system may be unstable if its angular momentum is taken away by magnetic torque only. The reason is that if there is a perturbation which increases the radial velocity, the inclination angle of the field will become smaller, which in turn increases the mass-loss rate and results in a higher radial velocity. Nevertheless, the linear stability analysis given by \citet{c2002b} suggested that a disk could be stable if the field is weak enough. \citet{l1994b} also stated that they didn't do a global calculation on the disk. Thus, a global time-dependent study should be necessary in order to investigate this problem, which is beyond the scope of this work. The formation of large scale
magnetic field also depends on the initial
strength and morphology of magnetic field as indicated by some
MHD simulations \citep*[e.g.,][]{i2003,b2008,m2012}, which is still
an open issue at present.

\section*{Acknowledgements}

We thank the referee for his/her very helpful report. We also thank M. C. Begelman, X. Cao and A. Tchekhovskoy for
helpful comments and discussion. This work is supported by the NSFC
(grants 11233006, 11373056) and the Science and Technology
Commission of Shanghai Municipality (10XD1405000).

\label{lastpage}

\end{document}